\documentclass[aps,pre,showpacs,reprint]{revtex4-1}

\usepackage{graphicx}% Include figure files
\usepackage{dcolumn}% Align table columns on decimal point
\usepackage{bm}% bold math

 \usepackage{color}

\usepackage{amssymb}
\usepackage{amsmath}
\usepackage{graphicx}
\usepackage{amsbsy}

\newcommand\beq{\begin{equation}}
\newcommand\eeq{\end{equation}}
\newcommand\beqa{\begin{eqnarray}}
\newcommand\eeqa{\end{eqnarray}}

\newcommand{\eff}{\phi_{\text{eff}}}

\newcommand{\phim}{\phi_{\text{J}}}

\newcommand{\phir}{\phi_{\text{rcp}}}

\begin{document}

%\preprint{}


\title{
A simple effective rule to estimate the jamming packing fraction of polydisperse hard spheres}


\author{Andr\'es Santos}
\email{andres@unex.es}
\homepage{http://www.unex.es/eweb/fisteor/andres/}
\author{Santos B. Yuste}
\email{santos@unex.es}
\homepage{http://www.unex.es/eweb/fisteor/santos/}
\affiliation{Departamento de F\'{\i}sica, Universidad de
Extremadura, Badajoz, E-06071, Spain}
\author{Mariano L\'{o}pez de Haro}
\email{malopez@unam.mx}
\homepage{http://xml.cie.unam.mx/xml/tc/ft/mlh/}
\affiliation{Instituto de Energ\'{\i}as Renovables, Universidad Nacional Aut\'onoma de M\'exico (U.N.A.M.),
Temixco, Morelos 62580, M{e}xico}
\author{Gerardo Odriozola}
\email{godriozo@imp.mx}
\affiliation{Programa de Ingenier\'{\i}a Molecular, Instituto Mexicano del Petr\'oleo, M\'exico, D. F. 07730, Mexico}
\author{Vitaliy Ogarko}
\email{v.ogarko@utwente.nl}
\affiliation{Multi Scale Mechanics (MSM), CTW, {MESA+,} University of Twente, PO Box 217, 7500 AE Enschede, The Netherlands}
\date{\today}

\begin{abstract}
A recent proposal in which the equation of state of a polydisperse hard-sphere mixture is mapped onto that of the one-component fluid is extrapolated beyond the freezing point to estimate the jamming packing fraction $\phi_\text{J}$ of the polydisperse system as  a simple function of $M_1M_3/M_2^2$, where $M_k$ is the $k$th moment of the size distribution. An analysis of experimental and simulation  data of $\phi_\text{J}$ for a large number of different mixtures shows a remarkable general agreement with the theoretical estimate.
To give extra support to the procedure, simulation data for  seventeen mixtures in the high-density region are used to infer the equation of state of the pure hard-sphere system in the metastable region. {An excellent} collapse of the inferred curves up to the glass transition and {a significant narrowing of the different out-of-equilibrium  glass branches all
the way to jamming are observed}.
Thus, the present approach  provides  an extremely simple  criterion to unify in a common framework and to give coherence to  data coming from very different polydisperse hard-sphere mixtures.
\end{abstract}

%\date{\today}

%\keywords{}

\maketitle
%\section{Introduction}
%\label{Intro}

Due to its simplicity and versatility, the hard-sphere (HS) system is considered as a prototype model in statistical physics \cite{M08}, being also  essential in studies in biological systems, granular matter, colloids, engineering, and materials science \cite{PZ10,TS10}. In {an equilibrium} monodisperse HS system the only control parameter is the packing fraction $\phi\equiv\frac{\pi}{6}\rho\sigma^3$ ($\rho$ and $\sigma$ being the number density and the diameter of the spheres, respectively), i.e., the volume occupied by the spheres, relative to the total volume $V$. It is well established that the equilibrium system is in  a stable fluid  phase from $\phi=0$ to the freezing packing fraction $\phi_\text{f}\simeq 0.492$ \cite{AW57,FMSV12}, undergoes a fluid-solid  transition  from $\phi_\text{f}$ to the  crystal melting point  $\phi_\text{m}\simeq 0.543$ \cite{HR68,FMSV12},  and finally is in a stable solid (crystalline) phase from $\phi_\text{m}$ to the close-packing fraction $\phi_{\text{cp}}=\frac{\pi}{6}\sqrt{2}\simeq 0.7405$, corresponding to face-centered-cubic close-packing \cite{S98,S98b}. However,
beyond the freezing point $\phi_{\text{f}}$ there is also a region of \emph{metastable} fluid states that  is supposed to end at the glass transition point  $\phi_\text{g} \simeq 0.58$ \cite{S98b,PZ10}. At even higher densities the system becomes a metastable amorphous solid  until it jams at a random close-packing {value} $\phi_{\text{rcp}}\simeq 0.64$.

{Although several semi-empirical equations of state have been proposed for the metastable fluid and glass regions \cite{H97b,MGPC08}, simulation studies in those regions} are particularly difficult \cite{RT96,S97,KLM04,BLW10,NCD13,PBC14} since one has to avoid the natural tendency to crystallization. Therefore, the understanding of the metastable properties of HS systems is still an active field of research and remains as an open problem. For instance, an important issue \cite{TTD00,KL07,WSJM11,JS13} is whether {the  concept of random close-packing is well defined}. {Indeed, as theoretically discussed in Ref.\ \cite{PZ10} and
numerically verified in Ref.\ \cite{CBS10}, the
jamming density depends on the fluid compression rate, even in the
absence of a crystal formation pathway}. Moreover, computer simulation results have shown that jamming can be achieved with an arbitrarily small increase of density at the expense of a correspondingly small increase in order \cite{TTD00,KL07}. This has led to the alternative definition of a ``maximally random jammed'' state, where a jammed packing is defined as a particle configuration in which each sphere is in contact with its nearest neighbors so that mechanical stability is conferred to the packing.

If this is so for monodisperse HSs, the situation becomes even more complicated for  {size-polydisperse} HS systems. Such systems have become recently of renewed interest and represent a very rich and active field of research {\cite{BCPZ09,FG09,ZVSPCP09,SW10,BCPZ10,KWP10,SVZPPC11,HJST11,HST12,OL12,OL13,HST13,SVZPCP14}}.
To begin with, it is well known that a certain degree of polydispersity inhibits the natural tendency of monodisperse HSs to  crystallize as the density is increased {\cite{P87b}}. Also, since by increasing polydispersity {the smaller spheres can either layer against larger spheres or fill the voids created between neighboring larger spheres, polydisperse HSs may pack to higher volume fractions than the monodisperse system {\cite{OL13}}.
Of course, in real materials the particle size distribution is {rarely} monodisperse,   so it is no surprise that polydisperse HSs have been  used  as models to study, among other subjects, storage in silos, metallic alloys, colloidal dispersions, solid propellants, and concrete.

As a consequence of their ability to attain higher packing fractions than the monodisperse system,  the jamming packing fraction $\phi_\text{J}$ of a polydisperse HS system is typically higher than the random close-packing fraction $\phir$ of the monodisperse system.
{Also, as happens with $\phir$, the value of $\phi_\text{J}$  depends on the out-of-equilibrium compression protocol followed to jam the system \cite{PZ10,CBS10}.}
Apart from that, since $\phim$ is a functional of the full size distribution {function} $f(\sigma)$, it differs widely  from system to system without an apparent unifying framework. It obviously would be desirable if it were possible to (i) characterize the whole distribution  by a single ``dispersity'' parameter $\lambda$ (with the convention that $\lambda=1$ defines the monodisperse case) and (ii) find a simple function $\phim(\lambda)$ such that the actual values of $\phim$ tend to approximately fall on the universal curve $\phim(\lambda)$. The main aim of this Rapid Communication is to show that this goal is indeed feasible,  $\lambda$ and $\phim(\lambda)$ being obtained  by extrapolating
{a simple model relationship between the equations of state of the equilibrium monodisperse and polydisperse fluids \cite{S12,S12c} to the metastable glass regime}.

The exact free energy of the mixture in a {spatially} uniform \emph{equilibrium} state must be consistent with two  independent ways of deriving the pressure $p$ \cite{R89,RELK02,R10} and with the  limit where one of the species is made of point particles \cite{S12}.
The fulfillment of these two exact conditions, together with the only  assumption (usual in fundamental-measure theories) that the excess free energy  depends only on $\rho$, $M_1$, $M_2$, and $M_3$ \cite{R89,SC98,W98,RELK02,R10,OL12}, with $M_k=\int_0^\infty d\sigma\,\sigma^k f(\sigma)$ being the $k$th moment of the size distribution, implies that the compressibility factor  $Z\equiv \beta{p}/{\rho}$ ($\beta$ being the  {inverse} temperature parameter) of the polydisperse HS mixture and the one ($Z_p$) of the pure HS system must \emph{necessarily} be related through \cite{S12c}
\beq
\phi Z(\phi)-\frac{\phi}{1-\phi}=\frac{m_3}{m_2^3}\left[\phi_{\text{eff}}Z_p(\phi_{\text{eff}})
-\frac{\phi_{\text{eff}}}{1-\phi_{\text{eff}}}\right] ,
\label{ZZnew}
\eeq
where $m_k\equiv M_k/M_1^k$ are (dimensionless) reduced moments and $\phi =\frac{\pi}{6}\rho M_3$ is the packing fraction of the polydisperse {system}.
{In Eq.\ \eqref{ZZnew}, $\eff$ is the \emph{effective} packing fraction the monodisperse fluid must have in order to serve as a reference to the state of the polydisperse fluid at {the} packing fraction $\phi$. Both fractions are related by}
\beq
\frac{\phi}{1-\phi}=\frac{m_3}{m_2^2}\frac{\phi_{\text{eff}}}{1-\phi_{\text{eff}}}.
\label{etaeffnew}
\eeq
{Equations \eqref{ZZnew} and \eqref{etaeffnew} have simple interpretations.
The quantity $\phi Z(\phi)-{\phi}/{(1-\phi)}$ represents the (reduced) excess pressure with respect to the ideal-gas value corresponding to the void volume $V(1-\phi)$. According to Eq.\ \eqref{ZZnew}, such an excess pressure in the mixture is simply proportional to the one in the effective one-component fluid (where the proportionality constant $m_3/m_2^3$ can be larger than, equal to, or smaller than {unity}). As for Eq.\ \eqref{etaeffnew}, it states that  the ratio between the occupied volume and the void one in the mixture is  larger by a factor $m_3/m_2^2$  than the same ratio on the effective pure fluid (note that $m_3\geq m_2^2\geq m_2\geq 1$ \cite{OL12}).
It is interesting to note that $\sqrt{m_3/m_2^2}=\sqrt{M_1M_3}/M_2$ represents (except for a factor $\sqrt{6\pi}$) the
ratio of the geometric mean of the average diameter and average volume of the spheres to their average area.}
The mapping defined by Eqs.\ \eqref{ZZnew} and \eqref{etaeffnew} is illustrated by Fig.\ \ref{fig0}.
The fact that the relationship between $Z$ and $Z_p$ is only established if they are evaluated at \emph{different} values of the packing fraction is in contrast with other approaches \cite{SYH11}, where both $Z$ and $Z_p$ are evaluated at the same $\phi$.

\begin{figure}[tbp]
  \includegraphics[width=8cm]{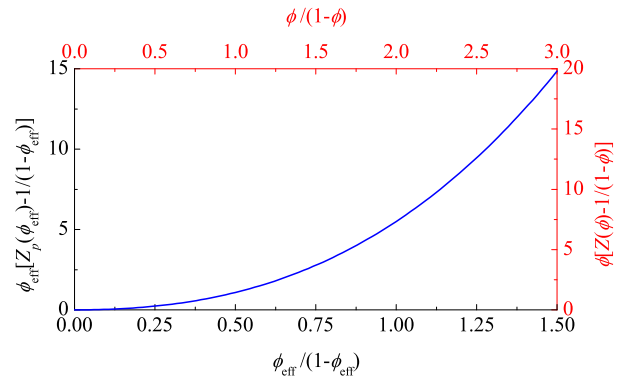}
\caption{(Color
  online) Graphical illustration of the mapping ``polydisperse mixture $\leftrightarrow$ pure fluid'' defined by Eqs.\ \protect\eqref{ZZnew} and \protect\eqref{etaeffnew}. A common curve represents the equation of state of the pure fluid (left and bottom axes) and that of the mixture (top and right axes). In this particular example the mixture is characterized by $m_2=\frac{3}{2}$ and $m_3=\frac{9}{2}$, so that $m_3/m_2^2=2$ and $m_3/m_2^3=\frac{4}{3}$. In this illustration the curve corresponds to the Carnahan--Starling \protect\cite{CS69} equation of state.} \label{fig0}
\end{figure}

Equation \eqref{ZZnew}  may be rewritten as
\beq
Z_p\left(\phi_\text{eff}\right)=m_2\left[Z(\phi)-\frac{1-m_2^{-1}}{1-\phi}\right]\left(1-\phi+\phi \frac{m_2^2}{m_3}\right).
\label{Zeff}
\eeq
{This implies that different polydisperse functions $Z(\phi)$ could be made to collapse {into} a common function $Z_p(\eff)$.}
Let us now  check whether  the results \eqref{ZZnew}--\eqref{Zeff}, originally derived for equilibrium fluid states,  are also (at least approximately) applicable to high-density {out-of-equilibrium} {metastable} states.
To that end, we performed simulations for the following classes of mixtures. First, we consider binary (B) mixtures  having a discrete composition characterized by
$f_\text{B}(\sigma)=(1-x)\delta(\sigma-a)+x\delta(\sigma-a w)$,
where $a$ and  $w$ are the small diameter and the ratio of the big to the small diameter, respectively, and $x$ is the mole fraction of the big species. Next, the top-hat (TH) distribution
$f_\text{TH}(\sigma)=\Theta(\sigma-a)\Theta(aw-\sigma)/a({w - 1})$, {where $\Theta$ is the Heaviside step function, has been chosen.} A generalization of the TH distribution is $f_\text{IP}(\sigma)=\Theta(\sigma-a)\Theta(aw-\sigma)\left(\sigma/a\right)^{-n}(n-1)/a\left(1 - w^{-n+1}\right)$, so that the distribution decays in the interval $a < \sigma < aw$ as an inverse power (IP) law of order $n$. Finally, the log-normal (LN) distribution is
$f_\text{LN}(\sigma)=\sigma^{-1}\exp\left[-\ln^2\left(\sigma/a\right)/2 s^2\right]/{\sqrt{2 \pi s^2}}$,
where $\ln a$ and $s$ are the average and the standard deviation, respectively,  of $\ln \sigma$.

\begin{table}
   \caption{Parameters of the mixtures considered in Fig.\ \protect\ref{fig1}.\label{table1}}
\begin{ruledtabular}
\begin{tabular}{lcccccc}
Label&$x$&$w$&$n$&$s$&{$m_3/m_2^2$}&{$m_3/m_2^3$} \\ \hline
B1&$0.2$&$1.3$&&&$1.0145$&$1.0016$\\
B2&$0.5$&$1.3$&&&$1.0162$&$0.9992$\\
B3&$0.2$&$1.4$&&&$1.0252$&$1.0032$\\
B4&$0.5$&$1.4$&&&$1.0256$&$0.9978$\\
B5&$0.2$&$1.6$&&&$1.0535$&$1.0073$\\
B6&$0.5$&$1.6$&&&$1.0454$&$0.9926$\\
TH1&&$5$&&&$1.0957$&$0.9543$\\
TH2&&$100$&&&$1.1249$&$0.8520$\\
IP1&&$1.4$&$2$&&$1.0094$&$0.9999$\\
IP2&&$10$&$2$&&$1.4071$&$0.9210$\\
IP3&&$20$&$2$&&$1.6555$&$0.8231$\\
IP4&&$2$&$3$&&$1.0407$&$1.0009$\\
IP5&&$5$&$3$&&$1.2354$&$1.0234$\\
IP6&&$10$&$3$&&$1.5278$&$1.0857$\\
LN1&&&&$0.1$&$1.0101$&$1$\\
LN2&&&&$0.5$&$1.2840$&$1$\\
LN3&&&&$0.7$&$1.6323$&$1$\\
     \end{tabular}
 \end{ruledtabular}
 \end{table}

\begin{figure}[tbp]
  \includegraphics[width=8cm]{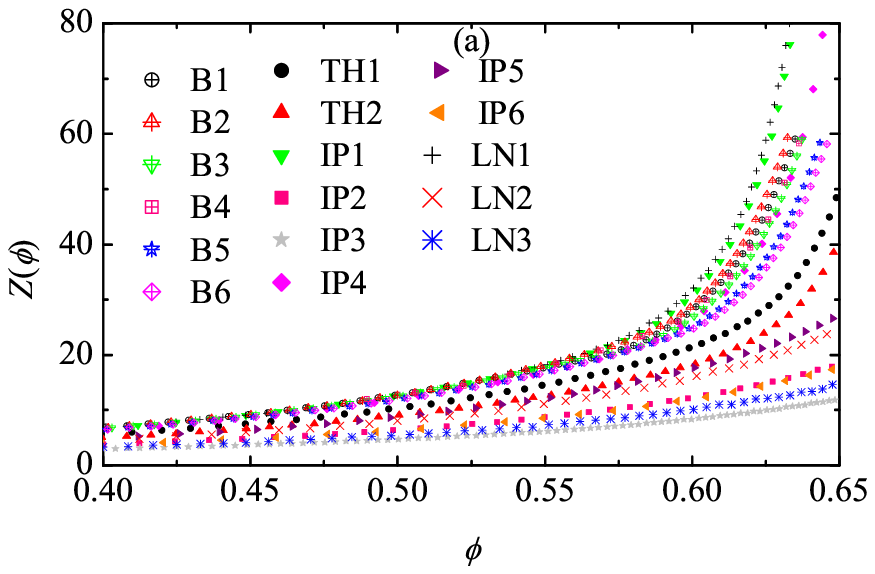}\\
  \includegraphics[width=8cm]{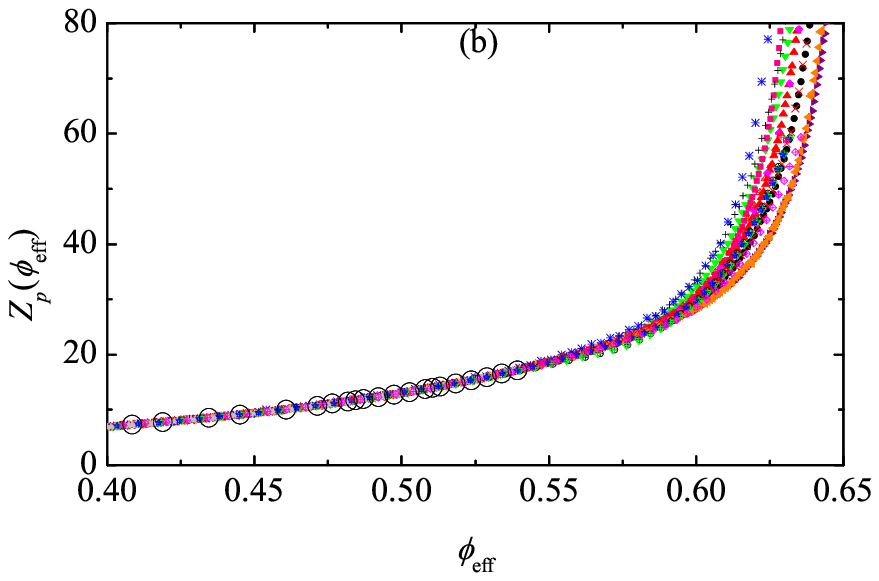}
\caption{(Color
  online) (a) Plot of $Z(\phi)$  for the six binary mixtures and eleven polydisperse mixtures considered in Table \protect\ref{table1}.
(b) Plot of the inferred one-component quantity $Z_p(\phi_\text{eff})$  associated with each one of the seventeen curves $Z(\phi)$ plotted in panel (a), as obtained from Eq.\ \protect\eqref{Zeff}. The circles represent simulation data for the true one-component system \protect\cite{KLM04,BLW10}.} \label{fig1}
\end{figure}

We have used the {replica exchange} Monte Carlo method {\cite{OKOM01,O09,OB11,note_14_04_1}} to obtain simulation data corresponding to six binary mixtures (B1--B6)  and event-driven molecular dynamics simulations, using a modification of the Lubachevsky--Stillinger algorithm {\cite{LS90,note_14_04_2}}, for eleven continuous distributions: two top-hat (TH1, TH2), three inverse-power with $n=2$ (IP1--IP3), three inverse-power with $n=3$ (IP4--IP6), and three log-normal (LN1--LN3). The parameters characterizing the seventeen different mixtures, together with the corresponding numerical values of the  {moment ratios $m_3/m_2^2$ and $m_3/m_2^3$},  are given in Table \ref{table1}.

Figure \ref{fig1}(a) displays the simulation results for the compressibility factor of those mixtures.
As expected, each mixture {widely} differs in the values of $Z$ for a common $\phi$. If the mapping ``polydisperse mixture $\leftrightarrow$ pure fluid'' characterized by Eqs.\ \eqref{ZZnew} and \eqref{etaeffnew} works, a high degree of collapse of the seventeen curves should be expected when the inferred one-component quantity $Z_p(\eff)$ [see Eq.\ \eqref{Zeff}]  is plotted instead of $Z(\phi)$. This is shown in Fig.\ \ref{fig1}(b), where we have also included the available simulation data for the {equilibrium and metastable} monodisperse HS system \cite{KLM04,BLW10}. We see that the collapse is almost perfect {for packing fractions below approximately  the glass transition value} $\phi_{\text{g}}\simeq 0.58$ \cite{PZ10,S98b},  well \emph{inside} the part of the metastable region where no data for the monodisperse system are available. {Beyond $\phi_\text{g}$, a certain degree of dispersion still remains, as expected from an algorithm-dependent out-of-equilibrium glass
branch. On the other hand, it is remarkable that the spread of the glass curves is certainly substantially smaller than the one in the {bare} mixture data} [cf.\ Fig.\ \ref{fig1}(a)]. This gives support to our expectation that the relationship implied by Eq.\ \eqref{ZZnew}  ({when extrapolated to} metastable states)  might be useful for inferring the equation of state of a metastable pure {HS} fluid from the knowledge of the high-density behavior of polydisperse HS mixtures, which is much more accessible than in the monodisperse case \cite{OB11}.

{Now we turn to our main point, namely the question of whether or not a single dispersity parameter $\lambda$ can be found as a tool to organize the apparently disconnected bunch of jamming packing values $\phim$ of different mixtures and, if so, what the relationship $\phim(\lambda)$ looks like. To that end, we push further the applicability of the model described by Eqs.\ \eqref{ZZnew} and \eqref{etaeffnew} and assume it is still valid (at least semiquantitatively) near the jamming point.} Since $\lim_{\phi \to \phi_\text{rcp}}Z_p(\phi)=\infty$ and $\lim_{\phi \to \phi_\text{J}}Z(\phi)=\infty$, Eq.\ \eqref{ZZnew} implies that $\phi_\text{J}$ is such that its associated value of $\phi_\text{eff}$ coincides with the random close-packing value $\phi_\text{rcp}$ of the pure system. Hence, from Eq.\ \eqref{etaeffnew} it follows that
\beq
\frac{\phi_{\text{J}}}{1-\phi_{\text{J}}}=\frac{m_3}{m_2^2}\frac{\phi_\text{rcp}}{1-\phi_\text{rcp}}.
\label{etaJ}
\eeq
{This equation fulfills the double requirement posed above. First, all the details of the size distribution {function} $f(\sigma)$ are encapsulated in the moment ratio $m_3/m_2^2$, which then plays the role of the sought dispersity parameter $\lambda$. Secondly, the functional form $\phim(\lambda)$ is quite simple: the occupied/void volume ratio at jamming, i.e., $\phi_{\text{J}}/(1-\phi_{\text{J}})$, is just \emph{proportional} to $\lambda$ with the slope being given by the one-component value ${\phi_\text{rcp}}/({1-\phi_\text{rcp}})$.}

Prediction \eqref{etaJ} is tested against experimental \cite{YCW65} and simulation \cite{BCPZ09,OL12,OL13,HST13,DW13} data  in Fig.\ \ref{fig3} taking $\phi_\text{rcp}=0.644$ \cite{BCPZ09}. As can be seen, although of course {an {unavoidable}  degree of scatter remains, the simple recipe \eqref{etaJ} satisfactorily succeeds in capturing the high correlation existing between $\phim$ and the dispersity parameter {$\lambda=m_3/m_2^2$} within the wide range $1\leq\lambda\lesssim 3$, thus giving order to  a large set of jamming densities corresponding to quite different  mixtures.  {The agreement is especially noteworthy in the case of experimental data for binary mixtures \cite{YCW65}}. The residual scatter of the data in Fig.\ \ref{fig3} reflects the influence on $\phim$ of   details of the size distribution not accounted for by $\lambda$ as well as of the  compression protocol followed in the simulations {\cite{OL12,OL13,CBS10}}.}

\begin{figure}[tbp]
  \includegraphics[width=8cm]{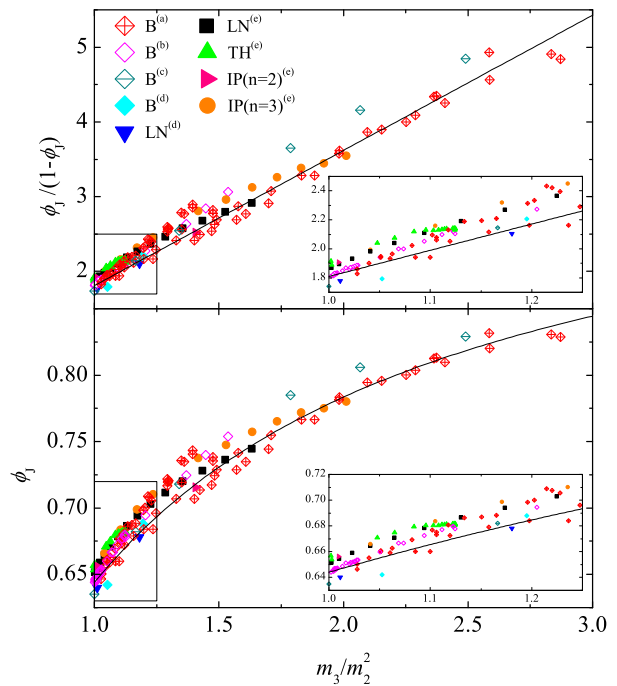}
\caption{(Color
  online) Test of the prediction for the jamming packing fraction of polydisperse HS mixtures [see Eq.\ \protect\eqref{etaJ}]. The labels (B, TH, IP, LN) denote the class of size distribution (see text). The superscripts indicate the sources: (a)  experimental data  from Ref.\ \protect\cite{YCW65}; (b) simulation data from Ref.\ \protect\cite{BCPZ09}; (c) simulation data from Ref.\ \protect\cite{HST13}; (d) simulation data from Ref.\  \protect\cite{DW13}; (e) simulation data from Refs.\ \protect\cite{OL12,OL13} and from this work. The insets are magnifications of the framed regions.
\label{fig3}}
\end{figure}

{Another strong argument in favor of the practical recipe \eqref{etaJ} is in order. In the case of a binary mixture at a fixed value $\eta\equiv(1-x)/(1-x+x w^3)$ of the volume fraction of small spheres, it is easy to prove that $m_3/m_2^2\to 1/\eta$ in the limit $w\to\infty$. Therefore,}
\beq
{\lim_{w\to\infty}\frac{\phi_{\text{J}}}{1-\phi_{\text{J}}}=\frac{1}{\eta}\frac{\phi_\text{rcp}}{1-\phi_\text{rcp}}.}
\label{etaJb}
\eeq
{It turns out that Eq.\ \eqref{etaJb} is an \emph{exact}  result \cite{YCW65,BCPZ10,FG09,KWP10} in the regime $\eta>(1-\phir)/(2-\phir)\simeq 0.26$ where all particles participate to the jammed structure (in contrast to the complementary regime where the small particles ``rattle'' in the voids made by the big particles). Thus, Eq.\ \eqref{etaJ} represents a generalization of the exact result \eqref{etaJb} to  binary mixtures with any size ratio $w$ and to mixtures with any size distribution {function} $f(\sigma)$.}

In summary, in the first part of this Rapid Communication  we have  used new simulation results to test the possibility of inferring the equation of state  of a metastable one-component HS fluid from data of a polydisperse HS mixture at high density {by means of Eqs.\ \eqref{ZZnew} and \eqref{etaeffnew}}. The main asset of this approach is that it is quite simple and yet leads to {an almost perfect agreement in the fluid regime and reasonably accurate predictions beyond the glass transition}. This compromise between simplicity and accuracy has been {proven} to be successful in the collapse that the curves of the different mixtures show  when the mapping ``polydisperse HS mixture $\leftrightarrow$ monocomponent HS fluid'' [cf.\ Eq.\ \eqref{Zeff}] is applied. This has
allowed  us to obtain educated estimates of pressure values  in a density range where, {due to technical difficulties, no simulation data for the HS monocomponent system seem to have been obtained so far.} {While  the inferred pressure curves beyond the glass point are algorithm-dependent, it can be speculated that a more reproducible out-of-equilibrium glass branch can be obtained in the limit of infinitely slow compression.} Secondly, and more importantly, the approach has been shown to  also be successful in the {approximate} prediction of the jamming packing fraction $\phi_\text{J}$  for polydisperse HS mixtures  as a function of the single parameter $m_3/m_2^2$
(i.e., the square of the ratio of the geometric mean of the average size and average volume of the spheres to their average area)
through the simple linear law \eqref{etaJ}.  Therefore, the present approach  provides  an extremely simple  criterion to organize and give coherence to  data coming from very different polydisperse HS mixtures.

%\begin{acknowledgments}
Three of the authors (A.S., S.B.Y., and M.L.H.) acknowledge the financial support from the Spanish Government and the Junta de Extremadura (Spain) through Grants No.\ FIS2010-16587 and  No.\ GR10158 (partially financed by FEDER funds), respectively. {The research of V.O. was supported
by the Dutch Technology Foundation STW (which is
the applied science division of NWO) and the Technology
Programme of the Ministry of Economic Affairs, Project No.\
STW-MUST 10120.} We also thank K. W. Desmond, E. R. Weeks, and F. Zamponi for making their simulation results available to us {and S. Luding for a critical reading of an early version of the manuscript}.

%\end{acknowledgments}

\bibliographystyle{apsrev}

\bibliography{D:/Dropbox/Public/bib_files/liquid}

\begin{thebibliography}{52}
\bibitem[{not({\natexlab{a}})}]{note_14_04_1}
\bibinfo{note}{The idea behind the replica exchange Monte Carlo method is to
  simulate several replicas of the same system at different but close enough
  thermodynamic states to allow efficient exchanges between the replicas. In
  the case of hard spheres, each replica samples an isobaric-isothermal
  ensemble at a different pressure, so that replicas having the larger
  pressures can escape from locally stable free energy minima through
  successive (detailed-balance) exchanges with replicas at lower pressures.}
\bibitem[{not({\natexlab{b}})}]{note_14_04_2}
\bibinfo{note}{In our molecular dynamics simulations the diameter $\sigma_i(t)$
  of particle $i$ was grown according to the linear law
  $\dot{\sigma}_i(t)=\Gamma \sqrt{2E/3M}\sigma_i(t)/\sigma_{\text{max}}(t)$,
  where $E$ and $M$ are the total energy and mass, respectively,
  $\sigma_{\text{max}}(t)$ is the largest diameter in the system, and the ratio
  $\sigma_i(t)/\sigma_{\text{max}}(t)$ is independent of time. We used slow
  growth rates $\Gamma=1.6\times 10^{-5}$--$1.6\times 10^{-4}$.}
\end{thebibliography}

\end{document}